\documentclass[journal]{IEEEtran}
\usepackage{amsmath,amsfonts}
\usepackage{graphicx}
\usepackage{epsf}
\usepackage{times}
\usepackage{float}
\usepackage{amssymb}
\usepackage{wasysym}
\usepackage[table]{xcolor} 
\usepackage{stfloats}
\usepackage{subeqnarray}
\usepackage[noadjust,compress]{cite}
\usepackage{multirow}
\usepackage{makecell}
\usepackage{booktabs}
\usepackage{siunitx}
\usepackage{pifont}
\usepackage{url}
\usepackage{hyperref}
\usepackage{tablefootnote}
\definecolor{red_cool}{rgb}{0.5, 0.0, 0.0}
\definecolor{tablecolor}{RGB}{255,240,240}
\definecolor{myblack}{rgb}{0,0,0}

\newcommand{\yuzhu}[1]{{\textcolor{myblack}{#1}}}

\begin{document}

\title{Moving Speaker Separation via Parallel Spectral-Spatial Processing}

\author{Yuzhu Wang, Archontis Politis, Konstantinos Drossos, Tuomas Virtanen
\thanks{
Signal Processing Research Center, Tampere University, Tampere, Finland (e-mail: yuzhu.wang@tuni.fi; archontis.politis@tuni.fi; tuomas.virtanen@tuni.fi). Nokia Technologies, Espoo, Finland (e-mail: konstantinos.drosos@nokia.com).
}
}

\markboth{IEEEtran \LaTeX \ Templates,~Vol.~18, No.~9, September~2020}%
{IEEEtran \LaTeX \ Templates}

\maketitle

\begin{abstract}
\label{sec:abstract}
Multi-channel speech separation in dynamic environments is challenging as time-varying spatial and spectral features evolve at different temporal scales. Existing methods typically employ sequential architectures, forcing a single network stream to simultaneously model both feature types, creating an inherent modeling conflict.
In this paper, we propose a dual-branch parallel spectral-spatial (PS2) architecture that separately processes spectral and spatial features through parallel streams. 
The spectral branch uses a bi-directional long short-term memory (BLSTM)-based frequency module, a Mamba-based temporal module, and a self-attention module to model spectral features.
The spatial branch employs bi-directional gated recurrent unit (BGRU) networks to process spatial features that encode the evolving geometric relationships between sources and microphones. Features from both branches are integrated through a cross-attention fusion mechanism that adaptively weights their contributions. 
Experimental results demonstrate that the PS2 outperforms existing state-of-the-art (SOTA) methods by 1.6-2.2 dB in scale-invariant signal-to-distortion ratio (SI-SDR) for moving speaker scenarios, with robust separation quality under different reverberation times (RT60), noise levels, and source movement speeds. Even with fast source movements, the proposed model maintains SI-SDR improvements of over 13 dB. These improvements are consistently observed across multiple datasets, including WHAMR! and our generated WSJ0-Demand-6ch-Move dataset.
\end{abstract}
%
\section{Introduction}
\label{sec:introduction}
\IEEEPARstart{M}{ulti-channel} speech separation aims to isolate individual speakers from their mixture signals captured by a microphone array. This process leverages both spectral and spatial cues~\cite{makino2007blind, benesty2008springer, vincent2018audio}.
Recent advances in this field have demonstrated potential for various applications, including voice assistants, hearing aids, and teleconferencing systems. 

Integration of deep neural networks (DNNs) into speech separation has driven significant advancements over the past decade~\cite{luo2019conv, tzinis2020sudo, luo2020dual, Chen2020DPTnet, Zeghidour2020, subakan2023exploring}. 
Multi-channel separation methods have been developed to operate on time-domain waveforms~\cite{luo2019fasnet, zhang2020end, zhang2021time} and time-frequency domain spectrograms~\cite{dai2024reference, Wang2023TFGridNet, Wang2023TFGridNet2, kalkhorani2024tf}.
Architecturally, researchers have developed both fully neural frameworks~\cite{Wang2023TFGridNet, Wang2023TFGridNet2, quan2024spatialnet, quan2024multichannel, kalkhorani2024tf} and hybrid systems that integrate DNNs with classical beamforming techniques~\cite{kubo2019mask, ochiai2023mask, wang2024attention, tammen2024array}. 
\begin{figure}
\centering
\includegraphics[width=0.97\columnwidth]{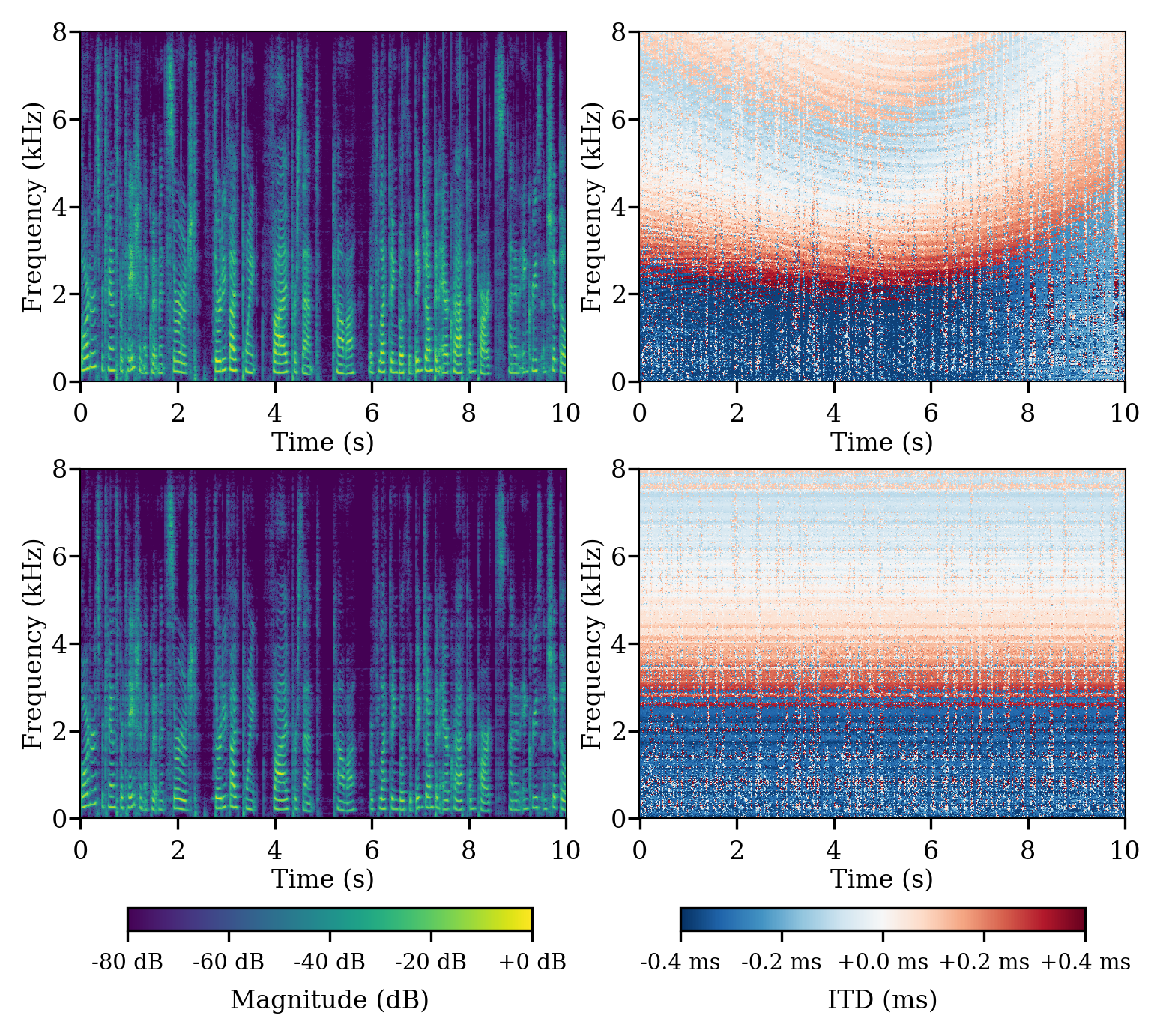}
\vspace{-2ex}
\caption{Visualization of temporal evolution in spectral and spatial features for moving versus static sound source scenarios. Top row: magnitude spectrogram (left) and inter-channel time difference (ITD) (right) of a moving speaker ($0.8$~m/s); Bottom row: corresponding visualizations for a static speaker positioned at the initial point of the moving source trajectory. The magnitude spectrograms use the first microphone channel of noise-free reverberant speech (RT$60$: $250$~ms) simulated in a $10$~m~$\times$~$10$~m~$\times$~$3$~m room with two microphones spaced by $15$~cm at $16$~kHz sampling rate.
The ITD is visualized by computing the time difference between the two channels for each time-frequency bin. For each channel, the complex-valued spectrogram is normalized by its magnitude, yielding a phase-only spectrogram. The inter-channel phase difference is then computed for each bin and transformed into a time delay.
}
\label{fig-scenario}
\vspace{-3ex}
\end{figure}

Despite the significant advancements in multi-channel speech separation, two main challenges remain unresolved. 
First, a key challenge in existing research is the common assumption that speakers remain static. In practice, speakers' spatial positions often change over time, resulting in time-varying propagation paths between the speakers and the microphone array. 
Second, another challenge lies in the sequential processing of spectral and spatial features through a single network processing stream. Most existing methods overlook the inherently different characteristics of these two feature types in speech signals.
As shown in Fig.~\ref{fig-scenario}, spectral and spatial features evolve on different temporal scales. When processed through a single network stream, the model must simultaneously handle two distinct processes with different temporal scales, potentially compromising its ability to effectively capture either aspect.

\begin{figure}
\centering
\includegraphics[width=0.95\columnwidth]{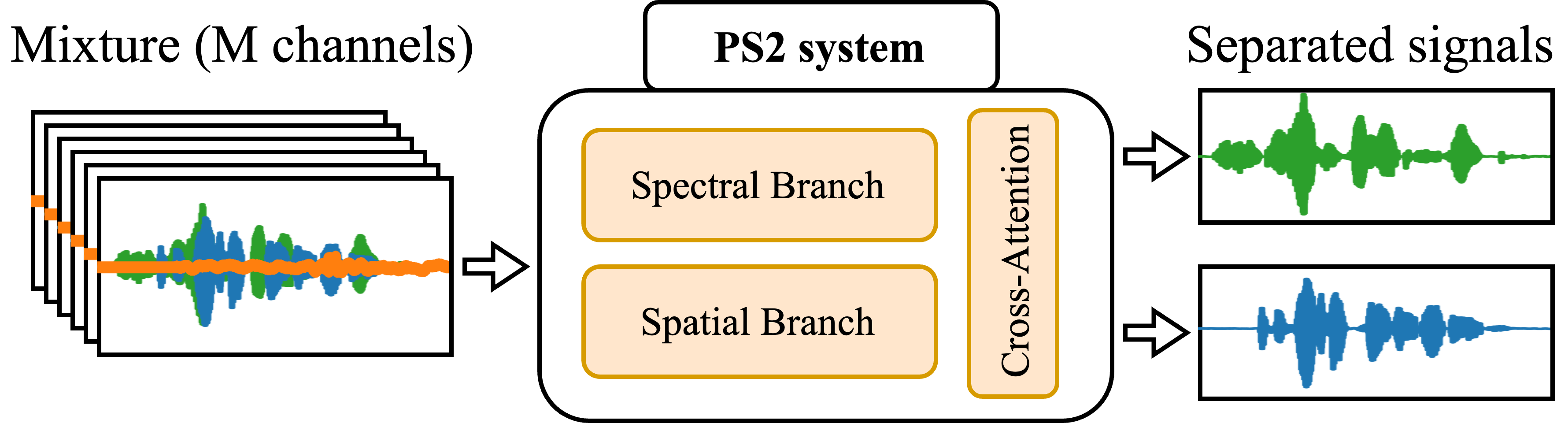}
\caption{The illustration of the proposed PS2 system.}
\label{fig-HL-system}
\vspace{-.5cm}
\end{figure}

To address these challenges, we propose a parallel spectral-spatial (PS2) architecture designed for multi-channel speech separation in dynamic scenarios. 
The key innovation of PS2 lies in its dual-branch design that processes input features in parallel, with one branch emphasizing spectral information and the other emphasizing spatial information, as shown in Fig.~\ref{fig-HL-system}.
Spectral features change rapidly over time, primarily influenced by speech production mechanisms~\cite{benesty2008springer}.
On the other hand, spatial features change more slowly over time, and the changes are over longer time scales~\cite{rossing2007springer, benesty2008springer}.
By dedicating separate processing streams to these two distinct feature types, our model better adapts to their different temporal scales. 

The spectral branch adopts the real-imaginary (RI) representation derived from the short-time Fourier transform (STFT) complex-valued spectrogram, following successful practices in the mapping of complex spectra~{\mbox{\cite{Wang2023TFGridNet, Wang2023TFGridNet2, quan2024spatialnet, kalkhorani2024tf}}}.
The spatial branch uses the magnitude-phase (MP) representation as input, which provides direct access to crucial spatial cues such as inter-channel phase and level differences (IPD and ILD, respectively)~\cite{adavanne2018sound, chakrabarty2019multi, adavanne2019localization, neri2024speaker}. 
To fuse the two processing streams, we develop a cross-attention module that modulates the contribution of spatial features based on their temporal coherence with spectral patterns, enabling adaptive feature integration responsive to varying acoustic conditions and source movement.

Our experimental results validate that the PS2 system outperforms existing methods across diverse acoustic conditions, with particular advantages in challenging scenarios involving rapid source movements.
Additionally, our sensitivity analysis demonstrates the effectiveness of the PS2 architecture in parallel feature processing.

The remainder of this paper is organized as follows: Section~\ref{sec:prior} reviews related works. Section III describes the signal model. Section IV introduces the PS2 system architecture. Section V describes the experimental setup. Section VI presents the experimental results and analysis, followed by conclusions in Section VII.
\section{Related works}
\label{sec:prior}
\subsection{DNN-based multi-channel source separation}
Deep learning has advanced multi-channel speech separation systems. Broadly, these systems can be classified into three categories. 
The first category extends single-channel separation systems to multi-channel ones~\cite{luo2019conv, tzinis2020sudo, luo2020dual, Chen2020DPTnet, Zeghidour2020, subakan2023exploring}. A typical strategy involves applying a single-channel separation system independently to each input channel, followed by combining learned features or masks to estimate the target signal~\cite{zhang2020end, zhang2021time, dai2024reference}. 

A second strategy is neural beamforming~\cite{kubo2019mask, ochiai2023mask, wang2024attention}, which combines DNNs with conventional beamforming techniques. 
A beamformer functions as a linear spatial filter designed by estimating or constraining spatial features in a mixture, including steering vectors, spatial covariance matrices, and IPD or ILD. 
Filter coefficients can be time-varying (adaptive beamforming) or time-invariant (fixed beamforming), with operations primarily conducted in the STFT domain. 
In neural beamforming, DNNs replace parameter estimation processes in the conventional beamforming framework.
DNNs' ability to integrate acoustic scene information and enable end-to-end optimization enhances beamforming, yielding improved parameter estimation and adaptability to complex environments.

The third category comprises fully neural multi-channel source separation systems that jointly learn from temporal, frequency, and channel dimensions~\cite{Wang2023TFGridNet, Wang2023TFGridNet2, quan2024spatialnet, quan2024multichannel, kalkhorani2024tf}.
TF-GridNet~\cite{Wang2023TFGridNet2} utilizes bi-directional long short-term memory (BLSTM) networks for temporal and frequency feature learning, while employing convolutional neural networks (CNNs) to encode channel information. 
SpatialNet~\cite{quan2024spatialnet} shares a similar architecture design with TF-GridNet but adopts Conformer blocks~\cite{quan2022conformer} for temporal feature extraction and convolutional-linear blocks for frequency feature extraction. Comparatively, TF-GridNet demonstrates strength in spectral information exploitation, while SpatialNet emphasizes spatial information processing.
\subsection{Moving source separation}
For moving source scenarios, various approaches have been proposed. In neural beamforming, attention mechanism~\cite{vaswani2017attention} enables tracking and enhancement of moving speakers through adaptive spatial filtering~\cite{ochiai2023mask, wang2024attention, tammen2024array}. Within blind source separation (BSS) framework, dynamic independent component/vector analysis methods have been developed to handle time-varying mixing conditions by updating separation matrices continuously~\cite{amor2021blind, koldovsky2021dynamic, jansky2022auxiliary}. Additional studies have integrated source localization with separation, using estimated directions-of-arrival (DOA) to guide the separation process~\cite{souden2013location, chang2019adaptive, taseska2017blind}.
However, moving source separation remains challenging due to several fundamental issues. Unlike static scenarios, source movement causes the acoustic transfer functions between sources and microphones to become time-varying, which significantly increases separation complexity and invalidates the time-invariance assumption that many existing algorithms rely on.
Additionally, DOA-based separation approaches suffer from degraded localization accuracy in noisy and reverberant environments. The DOA estimation error may increase dramatically during speech pauses, potentially causing separation system failure.
\subsection{Theoretical foundations of the dual-branch architecture}
Studies in speech production and acoustics have revealed that spectral and spatial features in speech signals evolve at different temporal scales and are influenced by different factors~\cite{deng2003analysis, deng2003analysischapter4, gannot2017consolidated, huang2008time, shaughnessy2008formant}.
As illustrated in Fig.~\ref{fig-scenario}, this temporal scale difference is observable when comparing the evolution patterns of spectral features and spatial cues in moving versus static source scenarios.
This fundamental difference provides a strong theoretical foundation for our PS2 system design.

Speech spectral features typically exhibit temporal scales that correspond to articulatory mechanisms. The fastest spectral transitions occur during stop consonant releases \mbox{($10$-$30$~ms)}, while formant transitions between vowels typically span $50$-$100$~ms~\cite{shaughnessy2008formant, deng2003analysis}. At the phoneme level, which represents the shortest meaningful units in speech, spectral variations occur within $50$-$150$~ms intervals~\cite{stevens2000acoustic, deng2003analysischapter7}. 

The evolution of spatial features follows different patterns. For static sources, while theoretically the spatial characteristics should remain constant, inevitable minor movements (e.g., head movements during speech) in real-world scenarios lead to gradual changes in relative transfer functions (RTFs). These variations typically occur at time scales ranging from hundreds of milliseconds to several seconds~\cite{gannot2017consolidated, deng2003analysischapter4, huang2008time}.
For moving sources, spatial feature dynamics become dependent on movement characteristics. 
Typical walking scenarios (about $1$~m/s) induce continuous spatial variations with significant changes observable over timescales of $100$-$200$~ms. Indoor activities, characterized by lower speeds but more complex trajectories, result in spatial variations that span a variable range of timescales, with significant changes observable from $100$~ms to several seconds~\cite{deng2003analysischapter4, gold2011speech}. In rapid movement scenarios, these changes can occur within tens of milliseconds.
\begin{figure}
\centering
\includegraphics[width=0.7\columnwidth]{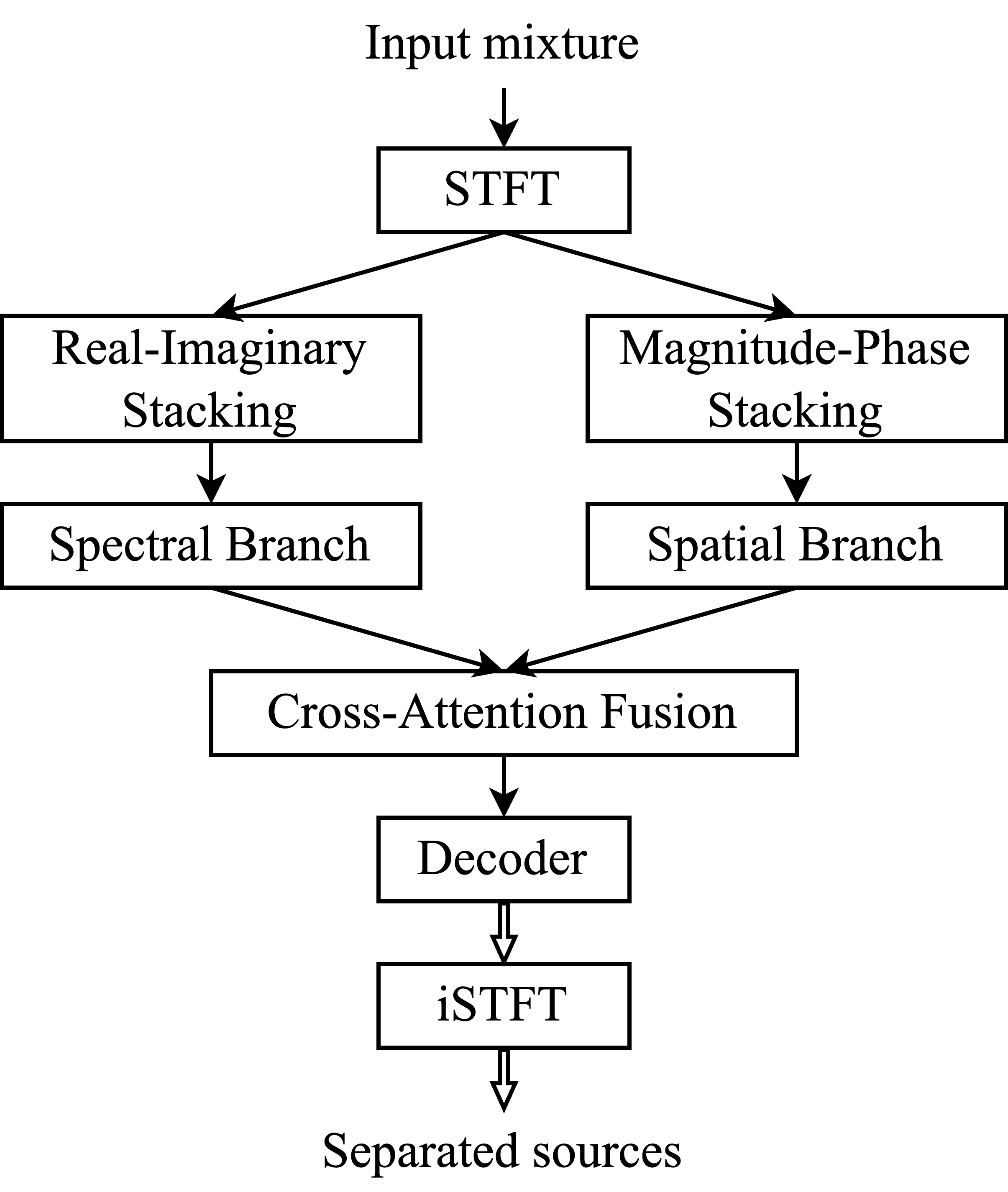}
\vspace{-1ex}
\caption{The architecture of the proposed PS2 system.}
\label{fig-architecture}
\vspace{-1ex}
\end{figure}
\section{Signal Model}
\label{sec:signal-model}
For $C$ speakers in a noisy environment, the captured signals by $M$ microphone channels can be represented in the time domain as
\begin{equation}
\label{eq:signal-model-time}
    {y}_m(n) = \sum_{c=1}^C {x}_{m,c}(n) + {v}_m(n), \ m\in\{1,...,M\}.
\end{equation}
${y}_m(n)$ represents the observed signal at the $m$-th microphone, ${x}_{m,c}(n)$ is the signal of speaker $c$ received by the $m$-th microphone, and ${v}_m(n)$ is the ambient noise at the $m$-th microphone.
Here, $m$, $c$, and $n$ denote microphone, speaker, and discrete-time indices, respectively. 

In the STFT domain, the signal model can be formulated as
\begin{equation}
\label{eq:signal-model-stft}
Y_{m}(t,f) = \sum_{c=1}^{C} X_{m,c}(t,f) + V_{m}(t,f),
\end{equation}
where $t$ and $f$ denote the STFT frame and frequency indices. 
$Y_{m}$, $X_{m,c}$, and $V_{m}$ represent the STFT of the observed signal, received speech, and noise, respectively.

In this study we operate under the assumptions that the number of sources $C$ is known and the microphone array geometry is fixed during both training and testing phases. 
Our objective is to separate $C$ signals $\{{x}_{m_{\text{ref}},c}(n)\}_{c=1}^C$ at the reference microphone $m_{\text{ref}}$ from the observed mixture signals $\{{y}_m(n)\}_{m=1}^M$, in scenarios involving either a single ($M = 1$) or multiple ($M > 1$) channels.

The separation task in moving source scenarios is more challenging than in static cases due to the time-varying nature of acoustic propagation. 
The signal of a static speaker at position $\mathbf{r}_s$ captured by a microphone at position $\mathbf{r}_m$ is modeled through the linear time-invariant system of 
\begin{equation}
    x_m(n) = s(n)*h(n|\mathbf{r}_s,\mathbf{r}_m) = \sum_{l=0}^L s(n-l)h(l|\mathbf{r}_s,\mathbf{r}_m),
\end{equation}
where $*$ denotes convolution. 
The room impulse response (RIR) is modeled as a finite impulse response (FIR) filter containing $L+1$ coefficients of non-negligible energy (as is typically the case with RIRs).
In the case of a moving speaker, the RIR at time $n$ from the speaker at position $\mathbf{r}_s(n)$ to the microphone is denoted as $h_n(l)$, hence it is time-dependent. The speech signal received at the microphone is then characterized by linear time-variant convolution as
\begin{equation}
    x_m(n) = \sum_{l=0}^L s(n-l)h_{n-l}(l|\mathbf{r}_s(n-l),\mathbf{r}_m).
\end{equation}
\section{PS2 system}
\label{sec:DynamicNet}
We propose PS2, an architecture that processes spectral and spatial features in parallel, as shown in Fig.~\ref{fig-architecture}. 
The input mixture signals are transformed using the STFT to produce complex-valued spectrograms $Y \in \mathbb{C}^{M \times T \times F}$, where $T$ is the number of frames and $F$ is the number of frequency bins.
The outputs from the spectral and spatial branches are fused through a cross-attention fusion module, which adaptively weights their contribution. The fused features are then processed by a decoder (2D deconvolution) to generate the predicted RI components for all the $C$ speakers. Finally, these separated spectral representations are transformed back to the time domain through the inverse STFT (iSTFT) to obtain the final separated speech signals.
We employ the scale-invariant signal-to-distortion ratio \mbox{(SI-SDR)}~\cite{LeRoux2018a} as the loss function during training.
\begin{figure}[tb]
\centering
\includegraphics[width=0.45\textwidth]{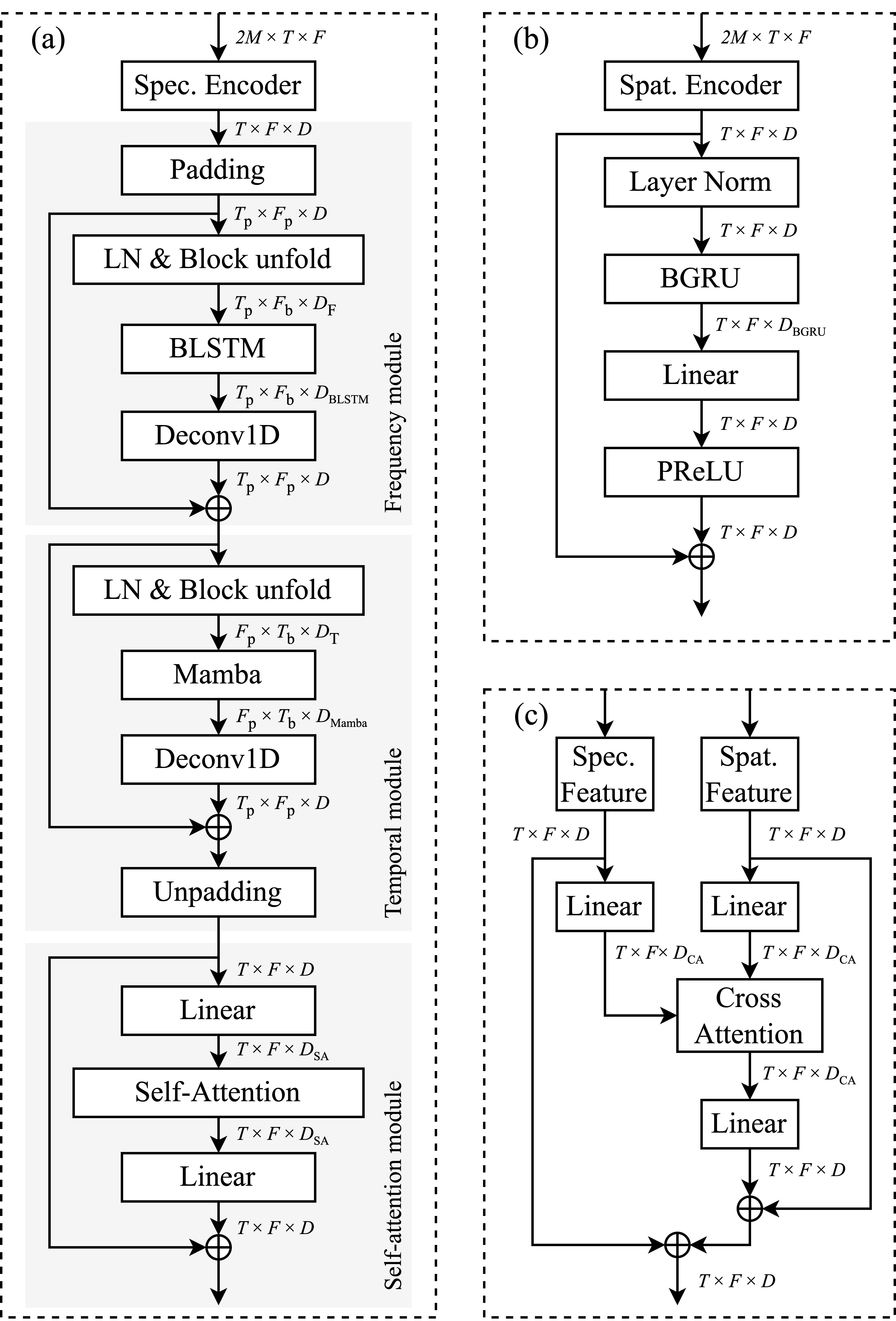}
\vspace{-1ex}
\caption{The core components of PS2 system: (a) spectral branch, (b) spatial branch, and (c) cross-attention fusion module.}
\label{fig-modules}
\vspace{-1ex}
\end{figure}
\subsection{Spectral Branch}
The spectral branch consists of four main components: the spectral branch encoder, the frequency module, the temporal module, and the self-attention module, as shown in Fig.~\ref{fig-modules}. 

The three modules in the spectral branch serve distinct functions: the frequency module performs intra-frame full-band processing (modeling all frequency bands within each time frame), the temporal module handles inter-frame sub-band modeling (processing temporal information within each frequency sub-band), and the self-attention module enables inter-frame full-band modeling (capturing temporal dependencies with all frequency bands across all time frames).

The input for the spectral branch is the stacked real and imaginary components $Y_{\text{RI}} \in \mathbb{R}^{2M \times T \times F}$. 
In the spectral branch encoder, we use a 2D convolutional layer across both time and frequency dimensions using a $3 \times 3$ kernel, followed by layer normalization. This produces a $D$-dimensional embedding for each T-F unit, yielding a tensor with shape $\mathbb{R}^{D \times T \times F}$. The tensor is then reshaped to $\mathbb{R}^{T \times F \times D}$.
We apply zero-padding to ensure proper block segmentation, resulting in a tensor of shape $\mathbb{R}^{T_{\text{p}}\times F_{\text{p}} \times D}$ with padded dimensions $T_{\text{p}}$ and $F_{\text{p}}$.
Subsequently, we perform block unfolding~\cite{Wang2023TFGridNet2} along the frequency dimension by segmenting it into overlapping blocks of size $I_{\text{F}}$ with a stride of $J_{\text{F}}$, yielding $F_{\text{b}}$ frequency blocks.
Here, block unfolding allows effective capture of correlations between adjacent frequency bands~\cite{Wang2023TFGridNet2}.
As a result, the features are reshaped to $\mathbb{R}^{T_{\text{p}}\times F_{\text{b}} \times D_{\text{F}}}$, where $D_{\text{F}} = D \cdot I_{\text{F}}$ represents the expanded feature dimension after concatenating features from $I_{\text{F}}$ adjacent frequency bands.

Then, we use a BLSTM network that operates along the frequency dimension, with $H_{\text{F}}$ hidden units in both directions (forward and backward).
The BLSTM processes each frame independently to capture intra-frame full-band dependencies.
After BLSTM, the dimensionality of the feature becomes $\mathbb{R}^{T_{\text{p}}\times F_{\text{b}} \times D_{\text{BLSTM}}}$, where $D_{\text{BLSTM}} = 2 \cdot H_{\text{F}}$ due to the concatenation of forward and backward hidden states.
Next, a one-dimensional deconvolution (Deconv1D) layer operating along the frequency dimension with kernel size $I_{\text{F}}$, stride $J_{\text{F}}$, input channel $D_{\text{BLSTM}}$ and output channel $D$ is applied to the outputs of the BLSTM.
Residual connections are employed between the input before the block unfolding and the output of the deconvolution layer, as shown in Fig.~\ref{fig-modules}.

In the temporal module, the processing pipeline is similar to that in the frequency module.
For temporal block unfolding, we segment the time dimension into overlapping blocks of size $I_{\text{T}}$ with a stride of $J_{\text{T}}$, yielding $T_{\text{b}}$ temporal blocks. 
The block unfolding allows modeling correlations between adjacent frames.
As a result, the features are reshaped to $\mathbb{R}^{F_{\text{p}}\times T_{\text{b}} \times D_{\text{T}}}$, where $D_{\text{T}} = D \cdot I_{\text{T}}$ represents the expanded feature dimension by concatenating features from $I_{\text{T}}$ adjacent time frames.
Along the temporal dimension, we employ a Mamba network~\cite{gu2023mamba} with a state dimension of $H_{\text{T}}$ to process the unfolded features, which produces output features with a shape of $\mathbb{R}^{F_{\text{p}}\times T_{\text{b}} \times D_{\text{Mamba}}}$. Here, Mamba is used because of its efficient sequential modeling capabilities.
A one-dimensional deconvolution (Deconv1D) layer operating along the temporal dimension with kernel size $I_{\text{T}}$, stride $J_{\text{T}}$, input channel $D_{\text{Mamba}}$ and output channel $D$ is applied.
This operation serves as the inverse of the block unfolding, restoring the block-wise features to their original temporal structure. Following this, the unpadding is performed to remove the padding previously added to both ends of the temporal and frequency dimensions.

In the self-attention module, a self-attention layer with $G_{\text{SA}}$ heads is employed along the temporal dimension to achieve inter-frame full-band processing, with residual connections applied as shown in Fig.~\ref{fig-modules}(a). 
To reduce the model's parameters, a linear layer is used before the self-attention layer to map the feature dimension $D$ into a lower-dimensional space $D_{\text{SA}}$. After the self-attention layer, another linear layer maps the features back to the original $D$ dimension. 
\subsection{Spatial Branch}
The spatial branch is designed to process spatial information of the acoustic scene, complementing the spectral modeling performed by the spectral branch.
The choice of magnitude-phase representation as input, inspired by localization techniques, facilitates direct access to spatial cues~\cite{chakrabarty2019multi}. 
The design of spatial branch was also inspired by the success of the convolutional recurrent neural network (CRNN) in sound event localization and detection (SELD) tasks~\cite{adavanne2018sound, adavanne2019localization, neri2024speaker}, where gated recurrent unit (GRU) network demonstrated superior performance.

The input for the spatial branch is the stacked magnitude and phase components $Y_{\text{MP}} \in \mathbb{R}^{2M \times T \times F}$.
The spatial branch encoder employs a 2D convolutional layer across both time and frequency dimensions with kernel size $3 \times 3$, followed by layer normalization. This operation transforms the input into a $D$-dimensional embedding for each T-F unit, producing a tensor with shape $\mathbb{R}^{D \times T \times F}$. As shown in Fig.~\ref{fig-modules}, the resulting tensor is reshaped to $\mathbb{R}^{T \times F \times D}$.
Then, we use a bi-directional gated recurrent unit (BGRU) network with $H_{\text{BGRU}}$ units in each direction.
We treat the features as a sequence of length $T$, with a feature dimension of $F \times D$.
A linear layer is applied to remap the feature dimensions $D_{\text{BGRU}}$ to the dimensions $D$, followed by a parametric rectified linear unit (PReLU) activation. 
Residual connections are used between the input of the BGRU and the output of PReLU.
\subsection{Cross-attention fusion module}
The cross-attention fusion module is designed to integrate outputs from the spectral branch and spatial branch. 
We explore two fusion approaches: sum-fusion (naive summation) and cross-attention fusion (weighted summation with learnable weights).
Sum-fusion directly sums the processed spectral and spatial features from their respective branches, offering simplicity but treating all features with equal and fixed weights across all time frames.
This fusion strategy is suboptimal for moving sound source scenarios, where the importance of spectral and spatial information varies continuously over time.
To address this limitation, we propose a cross-attention fusion module that dynamically adjusts the contribution of each branch based on its time-varying relative importance.

In the cross-attention fusion module, we first apply two independent linear layers to the spectral branch and spatial branch features, reducing their dimensionality from $D$ to $D_{\text{CA}}$ to decrease computational complexity. This results in two tensors of shape $\mathbb{R}^{T \times F \times D_{\text{CA}}}$.
We then reshape the reduced features to $\mathbb{R}^{T \times (F \times D_{\text{CA}})}$, treating each time frame as a sequence element with a feature dimension of $F \times D_{\text{CA}}$.
Following the standard multi-head attention mechanism~\cite{vaswani2017attention}, the reduced-dimension spectral branch features are linearly projected to form the query matrix $Q$, while the reduced-dimension spatial branch features are linearly projected to form the key $K$ and value $V$ matrices.
The cross-attention output is computed as
\begin{equation}
\mathrm{Attention}(Q,K,V) = \mathrm{softmax}\Bigl(\frac{QK^T}{\sqrt{D_{\text{CA}}}}\Bigr)V.
\end{equation}
As shown in Fig.~\ref{fig-modules}, the cross attention layer produces an output of shape $\mathbb{R}^{T \times F \times D_{\text{CA}}}$.
This output is then mapped back to the original dimension $\mathbb{R}^{T \times F \times D}$ through a linear layer. 
\subsection{Decoder}
The decoder employs a 2D deconvolution layer with $2C$ output channels and a $3 \times 3$ kernel to map the learned features from  $\mathbb{R}^{D \times T \times F}$ to $\mathbb{R}^{2C \times T \times F}$, where $C$ represents the number of sound sources. The $2C$ output channels predict the real and imaginary components for all $C$ speakers. 
\section{EXPERIMENTAL SETUP}
\label{sec:exp}
\subsection{Datasets}
\label{ssec:data}
For evaluating our model under noisy and reverberant conditions with static speakers, we employ the WHAMR! dataset ~\cite{Maciejewski2020WHAMR}. It builds upon the WSJ0-2mix dataset by adding reverberation to the clean signals and integrating non-stationary background noises to the mixtures. The dataset contains 20000 training samples, 5000 validation samples, and 3000 test samples.

Due to the lack of publicly available datasets for moving source separation, we synthesized a six-channel two-speaker dataset, WSJ0-Demand-6ch-Move. The speech signals were sourced from the WSJ0 corpus~\cite{paul1992design}, while the noise signals were extracted from the real-world noise recordings in the Demand dataset~\cite{thiemann2013diverse}. There was no overlap of speakers or noise signals in the generated training, validation, and test sets.

For each mixture signal, two utterances were randomly selected from WSJ0. We use the original level of speech signals in WSJ0. The relative energy ratio between any two utterances in WSJ0 is $[-5, 5]$~dB. The room dimensions were randomly generated for each mixture within the following ranges: length $[8, 10]$~m, width $[8, 10]$~m, and height $[3, 4]$~m. The reverberation time (RT60) of each room was randomly selected within the range $[0.1, 0.7]$~s.
We simulated a six-microphone circular array with the topology based on the microphone arrangement in the Demand noise dataset~\cite{thiemann2013diverse}.
We used microphones with indices $\{2, 3, 5, 7, 10, 11\}$ from the Demand array configuration~\cite{thiemann2013diverse}, and the array radius was $5$~cm.
The positions of the microphone array and the two sound sources were randomly generated for each mixture, where the microphone array and sound sources were constrained to be at least $0.5$~m away from any wall.
Additionally, the minimum distance between the microphone array and the sound sources, as well as between the two speakers, was set to $0.5$~m.
The height of the microphone array ranged from $[1.0, 1.5]$~m, while the height of the sound sources ranged from $[1.5, 2.0]$~m.
All microphones were placed at the same height above the ground.
Each sound source was assigned a random speed between $[0.0, 1.0]$~m/s. Based on this speed, a straight line motion trajectory was generated for each sound source. 

Given the generated room parameters and source trajectories, we used gpuRIR~\cite{diaz2021gpurir} to simulate six-channel reverberant speech signals for each sound source in the moving sound source scenario. The reverberant signals of the two sound sources were mixed in "min" mode (where the minimum length of the two signals determines the mixture length), followed by the addition of multichannel noise segments of equal length from the Demand dataset. Before mixing, the noise signals were scaled such that the signal-to-noise ratio (SNR) of the mixed signal varied within the range $[0, 10]$~dB. The SNR was calculated as the ratio between the average power of the two sources and the power of the noise signal.
We generated $25000$, $2500$, and $2500$ samples for the training, validation, and test sets, respectively. The generated signals have a $16$~kHz sampling frequency.

Using the same parameter settings, we also generated a six-channel two-speaker dataset with the same number of samples, WSJ0-Demand-6ch-Static, in which the spatial position of each sound source is the generated initial position.
\subsection{Training configuration}
\label{ssec:Training}
We used the ESPnet toolkit~\cite{watanabe2018espnet} for model training.
The Adam optimizer was employed with an initial learning rate of $5\times10^{-4}$.
Training was conducted for a maximum of $200$ epochs with an early stopping strategy and a batch size of $8$.
All hype-rparameters were determined through experiments on the validation set.

For the STFT at a $16$~kHz sampling rate, a window length of $512$ samples ($32$~ms) and a hop length of $256$ samples ($16$~ms) were used, employing a Hann window. 
The number of repeated blocks was set to $8$.  
The feature dimension $D$ was established at $48$. In the spectral branch, $I_{\text{F}}=3$, $J_{\text{F}}=1$, $I_{\text{T}}=3$, and $J_{\text{T}}=1$.

Within the frequency module in spectral branch, the BLSTM was configured with $H_{\text{F}}=96$ hidden units in each direction. The temporal module employed a Mamba architecture with a state feature dimension of $128$ and a convolutional feature dimension of $4$. 
The self-attention module was configured with $G_{\text{SA}}=4$ heads. 
The linear layer dimension $D_{\text{SA}}$ was $8$.
The spatial branch utilized a $2$-layer BGRU with a dropout rate of $0.05$ and $96$ hidden units in each direction.
In the cross-attention fusion module, the linear layer dimension $D_{\text{CA}}$ was set to $8$, with $4$ attention heads and a dropout rate of $0.05$.
Gradient clipping was implemented with an $L_2$ norm threshold of $5.0$. The training process did not incorporate dynamic mixing (DM) or data augmentation techniques.
\begin{table*}[t]
    \caption{\textsc{Ablation experiments.}}
    \vspace{-1.5ex}
    \label{tab-ablation}
    \centering
    \resizebox{0.95\textwidth}{!}{%
    \setlength{\tabcolsep}{6pt}
    \begin{tabular}{lccccccccccc}
    \toprule
    \textbf{\#} & Dual  &  Frequency  & Temporal & Self-attention & Spatial branch & Fusion & \textbf{\#}Params & FLOPs & SI-SDR  &  PESQ  & eSTOI   \\ 
     & branch &  module  & module & module & activation & method & (M) & (G/s) & (dB)  &  & \\
    \midrule
    Unproc. & -- & -- & -- & -- & -- & -- & -- & -- & -1.5 & 1.24 & 0.527 \\ 
    \midrule
    1 &  \checkmark &  BLSTM  & Mamba &  \checkmark & PReLU & Cross-attention &  8.4  & 66.8 &  \textbf{13.4} & \textbf{3.20}  & \textbf{0.897} \\ 
    2 &  \ding{55} & BLSTM  & Mamba&  \checkmark & -- & -- & 3.0 & 66.2 & 11.9 & 2.88 & 0.859 \\ 
    2+ &  \ding{55} & BLSTM  & Mamba&  \checkmark & -- & -- & 9.4 & 181.9 & 12.1 & 2.90 & 0.862 \\ 
    \midrule
    3 &  \checkmark & BLSTM  & Mamba &  \checkmark & Tanh & Cross-attention & 8.4 & 66.9 & 12.4  & 2.93 & 0.878  \\ 
    4 &  \checkmark & \ding{55}  & Mamba & \checkmark & PReLU & Cross-attention & 7.1 & 22.7 & 10.5  & 2.64 & 0.837 \\ 
    5 &  \checkmark & BLSTM  & \ding{55} & \checkmark & PReLU & Cross-attention & 7.3 & 46.3 & 10.9  & 2.72 & 0.841 \\
    6 &  \checkmark & BLSTM  & Mamba & \ding{55} & PReLU & Cross-attention & 7.8 & 65.1 & 12.2 & 2.87 & 0.879 \\
    7 &  \checkmark & BLSTM  & BLSTM & \checkmark & PReLU & Cross-attention & 8.6 & 90.5 & 13.2 & 3.14 & 0.891 \\
    8 &  \checkmark & Mamba  & BLSTM & \checkmark & PReLU & Cross-attention & 8.4 & 66.9 & 12.3 & 2.87 & 0.872 \\ 
    9 &  \checkmark & Mamba  & Mamba & \checkmark & PReLU & Cross-attention & 8.2 & 43.3 & 12.8 & 2.99 & 0.881 \\ 
    10 &  \checkmark & BLSTM  & Mamba & \checkmark & PReLU & Sum & 7.2 & 66.7 & 12.9 & 3.02  & 0.883 \\ 
    11 &  RI+MP & BLSTM  & Mamba & \checkmark & -- & -- & 3.0 & 66.2 & 11.8 & 2.81 & 0.855 \\ 
    12 &  RI/RI & BLSTM  & Mamba & \checkmark & PReLU & Cross-attention & 8.4 & 66.9 & 12.9 & 3.09 & 0.884 \\ 
    13 & \checkmark & BLSTM & BLSTM & \ding{55} & PReLU & Cross-attention & 7.9 & 88.9 & 12.1 & 2.85 & 0.878 \\
    14 & \checkmark & BLSTM & BMamba & \ding{55} & PReLU & Cross-attention & 9.1 & 85.9 & 12.2 & 2.88 & 0.879 \\
    \bottomrule
    \multicolumn{12}{p{0.95\textwidth}}{\textit{Notes}:}\\
    \multicolumn{12}{p{0.95\textwidth}}{(a) Deterministic mode was enabled in ESPnet~\cite{watanabe2018espnet}, which fixed the random seed for each epoch and the parameter initialization.}\\
    \multicolumn{12}{p{0.95\textwidth}}{(b) The results above are averages over multiple runs, with the observed run-to-run variability in SI-SDR being $\pm0.1$~dB (due to potential non-determinism in multi-GPU training and CuDNN/CUDA operations).}\\
    \multicolumn{12}{p{0.95\textwidth}}{(c) FLOPs are computed on 4-second segments and then divided by $4$ to yield values in Giga per second (G/s), where a 6-channel configuration with a batch size of $1$ and a sampling rate of $16$~kHz is used. We employ the \textit{torch.utils.flop\_counter} for the FLOPs calculation.}\\
    \end{tabular}
    }
    \vspace{-1.5em}
\end{table*}
\subsection{Evaluation metrics}
\label{ssec:metrics}
We employed objective metrics commonly used in the field to evaluate the proposed PS2 system in speaker separation tasks. These include the scale-dependent signal-to-distortion ratio (SDR)~\cite{vincent2006performance}, the scale-invariant signal-to-distortion ratio (SI-SDR)~\cite{LeRoux2018a}, source-to-interferences ratio (SIR)~\cite{vincent2006performance}, source-to-artifacts ratio (SAR)~\cite{vincent2006performance}, the perceptual evaluation of speech quality (PESQ)~\cite{rix2001perceptual}, and the extended version of the short-time objective intelligibility measure (eSTOI)~\cite{jensen2016algorithm}. For all these metrics, higher values indicate better performance. These metrics were calculated using the TorchMetrics library~\cite{detlefsen2022torchmetrics}.
\section{Evaluation results}
This section presents the separation performance of the PS2 system across various tasks and datasets.
\subsection{Ablation experiments}
\label{ssec:ablation}
To analyze the contribution of each component in the proposed PS2 system, we conduct a series of ablation experiments. These experiments disable or modify specific modules to analyze their impact on overall performance, using SI-SDR, PESQ, and eSTOI as evaluation metrics. The ablation experiments were conducted using the WSJ0-Demand-6ch-Move dataset. The results are summarized in Table~\ref{tab-ablation}. 

Experiment $1$ represents our proposed system with all components enabled: the dual-branch architecture processing both RI and MP representations, BLSTM for frequency modeling, Mamba for temporal modeling, self-attention modules for long-term modeling, PReLU as activation function in the spatial branch, and the cross-attention fusion module. This complete configuration achieves the best performance with an SI-SDR of $13.4$~dB, PESQ of $3.20$, and eSTOI of $0.897$, while maintaining reasonable computational complexity with $8.4$~M parameters and $66.8$~G/s FLOPs. This serves as our baseline for subsequent ablation studies.

Experiment $2$ examines the impact of removing the spatial branch, forcing the model to process both spectral and spatial information through a single sequential pipeline. This configuration results in a performance degradation, with SI-SDR dropping by $1.5$~dB (from $13.4$ to $11.9$~dB). This performance drop validates our core architectural hypothesis that separating the processing streams for spectral and spatial features is more effective than forcing a single network stream to simultaneously model both aspects. The increased modeling burden is evident in dynamic scenarios, where the network must simultaneously track rapid spatial changes while maintaining stable spectral feature extraction. This result aligns with our discussion in Section II, where we highlighted the fundamentally different characteristics of spectral and spatial features in speech signals, and supports our design choice of feature disentanglement through parallel processing streams.

To eliminate the potential influence of model size differences, we conducted experiment~$2$+, which maintains the single-branch design while increasing the feature dimension to $D=128$. This adjustment yields a model with a comparable parameter count to experiment $1$. Experiment $2$+ achieved an SI-SDR of $12.1$~dB, remaining $1.3$~dB lower than the dual-branch architecture in experiment $1$. This confirms that the performance degradation in experiment $2$ and $2$+ primarily stems from removing the spatial branch rather than from the change of parameter quantity.

In experiment $3$, we investigate the impact of activation functions in the spatial branch by replacing PReLU with Tanh. While Tanh is commonly used in GRU-based sound source localization tasks, our results show that PReLU achieves better performance in our dual-branch architecture, with SI-SDR improving by $1.0$~dB. 
This performance difference can be attributed to PReLU's learnable parameters, which provide additional flexibility in modeling the complex spatial relationships in our separation task, compared to the fixed non-linear mapping of Tanh.

Experiments $4$, $5$, and $6$ investigate the necessity of frequency modeling, temporal modeling, and self-attention modules in the spectral branch. Experiment $4$ disables the frequency modeling module while retaining temporal modeling and self-attention module, resulting in a substantial performance degradation with SI-SDR dropping by $2.9$~dB (from $13.4$ to $10.5$~dB).
Similarly, Experiment $5$ removes the temporal modeling module while keeping frequency modeling and the self-attention module, leading to significant performance drops with SI-SDR decreasing by $2.5$~dB (to $10.9$~dB). 
Experiment $6$ disables the self-attention module while maintaining frequency and temporal modeling, resulting in a performance reduction with SI-SDR dropping by $1.2$~dB (to $12.2$~dB). 

The frequency and temporal modeling modules are more critical than the self-attention module, as evidenced by the larger performance drops in Experiments $4$ and $5$ ($2.5$-$2.9$~dB) compared to Experiment $6$ ($1.2$~dB). The complementary nature of these modules is evident --- frequency modeling captures the spectral patterns characteristic of speech, temporal modeling tracks the evolution of these patterns over time, and self-attention enables long-term feature learning. The smaller impact of removing the self-attention module suggests that while it contributes to overall performance, the core separation capability is primarily driven by the frequency and temporal modeling components.

Experiments $7$, $8$, and $9$ investigate different architectural choices for frequency and temporal modeling modules, comparing BLSTM and Mamba implementations. Experiment $7$ employs BLSTM for both modules, achieving competitive performance (SI-SDR: $13.2$~dB)
but with significantly higher computational cost ($90.5$~G/s FLOPs). Experiment $8$ explores using Mamba for frequency modeling while retaining BLSTM for temporal modeling, resulting in reduced performance (SI-SDR: $12.3$~dB). Experiment $9$ implements Mamba for both modules, showing comparable performance (SI-SDR: $12.8$~dB) while achieving the lowest computational complexity ($43.3$~G/s FLOPs) among these variants. The experimental results demonstrate that both BLSTM and Mamba architectures effectively model frequency and temporal dependencies, with their optimal application determined by specific requirements. Notably, the Mamba-based implementation reduces computational complexity by up to $52\%$ compared to its BLSTM counterpart while maintaining comparable separation quality. The small performance variations across different variants ($0.2$-$1.1$~dB in SI-SDR) validate that the performance benefits of the dual-branch architecture stem primarily from the separation of spectral and spatial processing rather than specific network implementations.

Experiment $10$ compares different feature fusion strategies by replacing the cross-attention fusion mechanism with a simple summation. While the cross-attention approach achieves slightly better performance (SI-SDR: $13.4$~dB vs $12.9$~dB), the performance gap is relatively modest ($0.5$~dB in SI-SDR). The sum-fusion reduces the model parameters by $1.2$~M while maintaining competitive performance, suggesting it could be a viable alternative in scenarios where model size is prioritized over maximum separation quality.

Experiments $11$ and $12$ investigate different input configurations to understand the role of input representations. Experiment $11$ tests a single-branch architecture where RI and MP representations are stacked as input features to the spectral branch, while disabling the spatial branch. This configuration results in significantly degraded performance (SI-SDR: $11.8$~dB) compared to both the dual-branch baseline (Experiment $1$). 
\yuzhu{The similar performance between Experiments $11$ and $2$ (SI-SDR difference of $0.1$~dB) suggests that stacking RI and MP representations as input features does not effectively exploit their complementary characteristics within a single sequential processing stream.}
Experiment $12$ explores an alternative dual-branch configuration by replacing the MP representation in the spatial branch with RI representation. While this RI/RI dual-branch setup achieves competitive performance (SI-SDR: $12.9$~dB), it still falls short of the original RI/MP design by $0.5$~dB in SI-SDR. 
The observed performance gap between these configurations highlights the role of spectrum representation in network design. Despite the equivalence of RI and MP representations in terms of information content, their different realizations of complex-valued spectra provide implicit prior knowledge: the MP representation naturally emphasizes spatial cues through magnitude and phase differences, while the RI representation better preserves the complex-valued spectral structure.
The better performance of the RI/MP configuration demonstrates that appropriate feature representation can impact the model's performance, even when the underlying information content remains unchanged.

\yuzhu{Experiment $13$ employs BLSTM for both frequency and temporal modules without self-attention, achieving SI-SDR of $12.1$~dB. Comparing Experiments $7$ and $13$, the self-attention module contributes $1.1$~dB SI-SDR improvement, consistent with the $1.2$~dB contribution observed between Experiments $1$ and $6$. This consistency indicates that the self-attention module provides benefits across different choices of temporal modules. Although both self-attention and temporal modules operate along the temporal dimension, the contribution of self-attention suggests that their modeling mechanisms are complementary rather than redundant.
Experiment $14$ employs bidirectional Mamba (BMamba)\footnote{\href{https://github.com/JusperLee/SPMamba}{github.com/JusperLee/SPMamba}}~\cite{li2024spmamba} for temporal module, achieving SI-SDR of $12.2$~dB with $85.9$~G/s FLOPs. Comparing Experiments $6$ and $14$, Mamba and BMamba exhibit comparable modeling capability, while Mamba uses fewer parameters ($7.8$~M vs $9.1$~M) and offers superior computational efficiency ($65.1$~G/s vs $85.9$~G/s FLOPs). This is the reason we adopt Mamba rather than BMamba in this study. Similar findings have also been reported in recent studies~\cite{li2024spmamba, zhu2024vision, jiang2025dual}.}

Besides, we experimented with swapping the order of the frequency and temporal modules in the spectral branch but found that the results showed almost no difference.
\subsection{Results on WSJ0-Demand-6ch-Move}
We evaluate the models' ability to handle dynamic scenarios on the WSJ0-Demand-6ch-Move, as shown in Table~\ref{table_moving_MC_SC}, where speaker positions change throughout the utterances. This evaluation is challenging as it requires models to track and separate moving sources while maintaining separation quality.

In the multi-channel setting, TF-GridNet achieves moderate results in dynamic conditions, with SI-SDR of $11.8$~dB.
SpatialNet shows varying performance across its configurations.
The small version ($1.6$~M parameters) achieves SI-SDR of $10.4$~dB, while the large version ($7.3$~M parameters) yields SI-SDR of $12.1$~dB.
However, both versions show lower performance compared to their static-scenario results in Table~\ref{table_static_MC_SC}, with drops of $3.8$~dB and $3.5$~dB in SI-SDR respectively, suggesting challenges in modeling dynamic conditions.
PS2 system outperforms both TF-GridNet and SpatialNet in moving source scenarios (SI-SDR: $13.4$~dB), showing improvements of $1.6$~dB in SI-SDR over TF-GridNet and $2.2$~dB over SpatialNet(large). The performance degradation from static to dynamic conditions is also smaller ($2.8$~dB in SI-SDR).

Table~\ref{table_moving_MC_SC} also presents single-channel evaluation results. 
TF-GridNet maintains moderate performance (SI-SDR: $8.8$~dB), showing a drop of $3$~dB in SI-SDR from its multi-channel results. SpatialNet exhibits more substantial degradation, particularly in its small version (SI-SDR drop of $3.7$~dB to $6.7$~dB). 
The PS2 achieves SI-SDR of $10.3$~dB in single-channel scenarios, representing a $3.1$~dB drop from the multi-channel performance.

The experimental results demonstrate the advantages of separating spectral and spatial processing streams in dynamic scenarios. Notably, the PS2 achieves substantial performance improvements compared to TF-GridNet and SpatialNet in single-channel scenarios, where spatial cues are unavailable. This advantage can be explained by the dual-branch design of the PS2, which allows separate modeling of time-varying acoustic transfer functions caused by source movement and the inherent spectral variations in speech signals. Although in single-channel settings, movement-induced transfer function changes are embedded within spectral variations, they still exhibit different temporal scales compared to speech spectral features, making the dual-branch processing beneficial. When spatial information becomes available in multi-channel settings, the experimental results show further performance enhancement.
\subsection{Results on WSJ0-Demand-6ch-Static}
We evaluate separation performance on the WSJ0-Demand-6ch-Static dataset under both multi-channel and single-channel conditions, as shown in Table~\ref{table_static_MC_SC}. This evaluation represents the common setting where speaker positions remain fixed throughout the utterances, which serves as a standard benchmark in speech separation research.

\begin{table}[tpb]
\setlength\tabcolsep{2pt}
\renewcommand{\arraystretch}{1.0}
\caption{\textsc{Results on WSJ0-Demand-6ch-Move dataset under multi-channel and single-channel conditions.}}
\vspace{-1.5ex}
\label{table_moving_MC_SC}
\centering
\resizebox{\linewidth}{!}{
\begin{tabular}{lcccccccc}
\toprule
\multirow{2}{*}{Systems} & \#Param & \multirow{2}{*}{Config.} & SDR & SI-SDR & SIR & SAR & PESQ & eSTOI  \\
 & (M) & & (dB) & (dB) & (dB) & (dB) &  &  \\
\midrule
\multirow{2}{*}{Unproc.} & \multirow{2}{*}{-} & Multi & -1.4 & -1.5 & 0.1 & 7.6 & 1.24 & 0.358 \\
 & & Single & -1.5 & -1.6 & 0.1 & 7.5 & 1.21 & 0.347 \\
\midrule
\multirow{2}{*}{TF-GridNet$^{\text{2}}$~\cite{Wang2023TFGridNet2}} & \multirow{2}{*}{14.5} & Multi & 12.1 & 11.8 & 23.1 & 12.6 & 2.85 & 0.857 \\
 & & Single & 9.2 & 8.8 & 19.5 & 9.8 & 2.32 & 0.752 \\
\midrule
\multirow{2}{*}{SpatialNet(small)$^{\text{3}}$~\cite{quan2024spatialnet}} & \multirow{2}{*}{1.6} & Multi & 10.8 & 10.4 & 19.7 & 11.7 & 2.52 & 0.815 \\
 & & Single & 7.2 & 6.7 & 17.1 & 7.9 & 2.05 & 0.695 \\
\midrule
\multirow{2}{*}{SpatialNet(large)$^{\text{3}}$~\cite{quan2024spatialnet}} & \multirow{2}{*}{7.3} & Multi & 11.6 & 11.2 & 21.2 & 12.3 & 2.73 & 0.839 \\
 & & Single & 8.1 & 7.6 & 18.3 & 8.7 & 2.21 & 0.728 \\
\midrule
\multirow{2}{*}{PS2} & \multirow{2}{*}{8.4} & Multi & \textbf{13.7} & \textbf{13.4} & \textbf{25.5} & \textbf{14.1} & \textbf{3.20} & \textbf{0.897} \\
 & & Single & \textbf{10.9} & \textbf{10.3} & \textbf{21.2} & \textbf{11.4} & \textbf{2.65} & \textbf{0.811} \\
\bottomrule
\end{tabular}
}
\vspace{-0.4cm}
\end{table}
\begin{table}[tpb]
\setlength\tabcolsep{2pt}
\renewcommand{\arraystretch}{1.0}
\caption{\textsc{Results on WSJ0-Demand-6ch-Static dataset under multi-channel and single-channel conditions.}}
\vspace{-1.5ex}
\label{table_static_MC_SC}
\centering
\resizebox{\linewidth}{!}{
\begin{tabular}{lcccccccc}
\toprule
\multirow{2}{*}{Systems} & \#Param & \multirow{2}{*}{Config.} & SDR & SI-SDR & SIR & SAR & PESQ & eSTOI  \\
 & (M) & & (dB) & (dB) & (dB) & (dB) &  &  \\
\midrule
\multirow{2}{*}{Unproc.} & \multirow{2}{*}{-} & Multi & -1.4 & -1.5 & 0.1 & 7.6 & 1.24 & 0.358 \\
 & & Single & -1.5 & -1.6 & 0.1 & 7.5 & 1.21 & 0.347 \\
\midrule
\multirow{2}{*}{TF-GridNet\tablefootnote{\href{https://github.com/espnet/espnet/blob/master/espnet2/enh/separator/tfgridnet_separator.py}{espnet2/enh/separator/tfgridnet\_separator.py}}~\cite{Wang2023TFGridNet2}} & \multirow{2}{*}{14.5} & Multi & 14.8 & 14.4 & 27.8 & 15.3 & 3.21 & 0.909 \\
 & & Single & 11.1 & 10.9 & 23.3 & 11.7 & 2.78 & 0.842 \\
\midrule
\multirow{2}{*}{SpatialNet(small)\tablefootnote{\url{https://github.com/Audio-WestlakeU/NBSS}}~\cite{quan2024spatialnet}} & \multirow{2}{*}{1.6} & Multi & 14.6 & 14.2 & 26.3 & 14.9 & 3.13 & 0.896 \\
 & & Single & 9.2 & 8.8 & 21.4 & 9.8 & 2.48 & 0.795 \\
\midrule
\multirow{2}{*}{SpatialNet(large)$^{\text{3}}$~\cite{quan2024spatialnet}} & \multirow{2}{*}{7.3} & Multi & 15.1 & 14.7 & 28.9 & 15.7 & 3.24 & 0.912 \\
 & & Single & 10.1 & 9.6 & 22.7 & 10.7 & 2.65 & 0.823 \\
\midrule
\multirow{2}{*}{PS2} & \multirow{2}{*}{8.4} & Multi & \textbf{16.4} & \textbf{16.2} & \textbf{31.9} & \textbf{16.9} & \textbf{3.47} & \textbf{0.938} \\
 & & Single & \textbf{13.4} & \textbf{13.1} & \textbf{25.2} & \textbf{12.9} & \textbf{3.02} & \textbf{0.882} \\
\bottomrule
\end{tabular}
}
\vspace{-0.4cm}
\end{table}

In the multi-channel setting, all models demonstrate substantial improvements over the unprocessed signals. TF-GridNet achieves strong performance (SI-SDR: $14.4$~dB). Notably, SpatialNet exhibits efficient spatial information utilization, with its small version ($1.6$~M parameters) achieving competitive performance (SI-SDR: $14.2$~dB). The large version of SpatialNet outperforms TF-GridNet (SI-SDR: $14.7$~dB) while using only half the parameters ($7.3$~M vs $14.5$~M). This efficiency can be attributed to SpatialNet's extensive use of CNN-based architectures, where parameter-sharing mechanisms reduce model size while maintaining strong spatial modeling capabilities.
The PS2 system achieves higher performance in the multi-channel scenario (SI-SDR: $16.2$~dB), showing improvements of $1.5$~dB in SI-SDR over SpatialNet(large). The performance advantage suggests that explicitly separating spectral and spatial processing streams benefits separation quality even in static conditions.

Table~\ref{table_static_MC_SC} also presents evaluation results for single-channel separation, where only the first channel of the WSJ0-Demand-6ch-Static dataset is used. When evaluating single-channel performance, the models exhibit different levels of robustness to the lack of spatial information. TF-GridNet maintains reasonable performance (SI-SDR: $10.9$~dB), with a performance drop of $3.5$~dB in SI-SDR. 
In contrast, SpatialNet shows more substantial degradation (SI-SDR drop of $5.1$~dB for the large version, from $14.7$~dB to $9.6$~dB), particularly in its small version (SI-SDR drop of $5.4$~dB, from $14.2$~dB to $8.8$~dB). This larger performance drop aligns with SpatialNet's architectural focus on spatial feature modeling and confirms its strong reliance on spatial cues.
The PS2 system demonstrates better resilience to the loss of spatial information, achieving the best single-channel performance (SI-SDR: $13.1$~dB) with the smallest performance drop ($3.1$~dB in SI-SDR). 

These results suggest that the dual-branch architecture of the PS2 system offers advantages in both multi-channel and single-channel scenarios, while maintaining relatively consistent performance across different channel configurations. The analysis reveals the trade-offs between model complexity, spatial information utilization, and robustness to channel reduction across different architectural designs.
\begin{table}[tpb]
\setlength\tabcolsep{2pt}
\renewcommand{\arraystretch}{1.0}
\caption{\textsc{Results on WHAMR! (1ch).}}
\vspace{-1.5ex}
\centering
\begin{tabular}{
p{3.5cm}
c
c
c
c
}
\toprule
Systems & {SDR(dB)} & {SI-SDR(dB)} & {PESQ} & {ESTOI}  \\
\midrule
unproc. & -3.5 & -6.1 & 1.41 & 0.317 \\ %
\midrule
Conv-TasNet~\cite{luo2019conv, Maciejewski2020WHAMR} & -- & 2.2 & -- & -- \\
BLSTM-TasNet~\cite{Maciejewski2020WHAMR} & -- & 3.0 & -- & -- \\
SISO$_1$~\cite{Wang2020css} & 6.2 & 4.2 & 1.79 & 0.594 \\
DPRNN~\cite{luo2020dual} & -- & 4.2 & -- & -- \\
DPTNET~\cite{Chen2020DPTnet} & 7.6 & 6.0 & -- & -- \\
Gated DPRNN~\cite{Nachmani2020} & -- & 6.1 & -- & -- \\
Wavesplit + DM~\cite{Zeghidour2020} & 8.7 & 7.1 & -- & -- \\
Sepformer + DM~\cite{subakan2023exploring} & 9.5 & 7.9 & -- & -- \\
SUDO RM -RF + DM~\cite{tzinis2020sudo} & -- & 7.4 & -- & -- \\
QDPN + DM~\cite{Rixen2022QDPN} & -- & 8.3 & -- & -- \\
MossFormer + DM~\cite{zhao2023mossformer} & -- & 10.2 & -- & -- \\
DasFormer~\cite{wang2023dasformer} & -- & 11.2 & -- & -- \\
TF-GridNet~\cite{Wang2023TFGridNet2} & 12.3 & 11.2 & 2.79 & 0.808 \\
SpatialNet(large)~\cite{quan2024spatialnet} & 11.2 & 10.2 & 2.54 & 0.772 \\
TF-CrossNet~\cite{kalkhorani2024tf} & 13.1 & 12.0 & 2.91 & 0.824 \\
\midrule
PS2 & 12.5 & 11.4 & \textbf{2.95} & \textbf{0.831} \\
\bottomrule
\end{tabular}
\vspace{-0.4cm}
\label{results_single_whamr}
\end{table}
\begin{table}[tpb]
\setlength\tabcolsep{2pt}
\renewcommand{\arraystretch}{1.0}
\caption{\textsc{Results on WHAMR! (2ch).}}
\vspace{-1.5ex}
\centering
\begin{tabular}{
p{3.5cm}
c
c
c
c
}
\toprule
Systems & {SDR(dB)} & {SI-SDR(dB)} & {PESQ} & {ESTOI}  \\
\midrule
unproc. & -3.5 & -6.1 & 1.41 & 0.317 \\ %
\midrule
MC-ConvTasNet~\cite{zhang2020end, zhang2021time} & -- & 6.0 & -- & -- \\
\makecell[l]{MC-ConvTasNet with\\speaker extraction~\cite{zhang2021time}} & -- & 7.3 & -- & -- \\
TF-GridNet~\cite{Wang2023TFGridNet2} & 14.8 & 13.7 & 3.16 & 0.868 \\
SpatialNet(small)~\cite{quan2024spatialnet} & 13.1 & 11.8 & 2.93 & 0.826 \\
SpatialNet(large)~\cite{quan2024spatialnet} & 15.0 & 14.1 & 3.16 & 0.870 \\
\midrule
PS2 & 14.9 & 13.9 & \textbf{3.21} & \textbf{0.881} \\
\bottomrule
\end{tabular}
\vspace{-0.4cm}
\label{results_whamr}
\end{table}
\subsection{Results on WHAMR!}
We evaluate the PS2 system on both single-channel and multi-channel configurations of the WHAMR! dataset, where noisy and reverberant mixtures are used as input, and the target is dry (anechoic and noise-free) speech signals. The evaluation setup creates a joint task of separation, dereverberation, and denoising.
Unless otherwise specified, we employ the complete model architecture (experiment \#$1$ in Table~\ref{tab-ablation}) for all subsequent experiments. 

For single-channel evaluation shown in Table~\ref{results_single_whamr}, we compare the PS2 system with various existing approaches. 
\yuzhu{Methods marked with "$+$DM" employed dynamic mixing for data augmentation during training.}
The PS2 system achieves an SI-SDR of $11.4$~dB, which is slightly lower than TF-CrossNet ($12.0$~dB) and outperforms previous SOTA methods such as TF-GridNet ($11.2$~dB) and DasFormer ($11.2$~dB).
Notably, the PS2 system demonstrates superior speech quality metrics (PESQ: $2.95$, ESTOI: $0.831$), suggesting effective preservation of speech characteristics during separation.

In the multi-channel setting shown in Table~\ref{results_whamr}, the PS2 system achieves strong performance (SI-SDR: $13.9$~dB) comparable to TF-GridNet and SpatialNet(large). Furthermore, the PS2 system achieves speech quality metrics (PESQ: $3.21$, ESTOI: $0.881$) that slightly exceed current SOTA approaches. The performance gap between single-channel and multi-channel configurations (approximately $2.5$~dB in SI-SDR) indicates the contribution of spatial information when available.
\subsection{Performance Analysis under Different Acoustic Conditions}
To gain deeper insights into model behavior under various conditions, we analyze the performance of our trained model on different subsets of test data. While the model is trained on the complete WSJ0-Demand-6ch-Move dataset, we partition the test set based on specific acoustic parameters to examine how separation performance varies with different conditions.

\begin{figure}[!tbp]
  \centering
  \includegraphics[width=89mm]{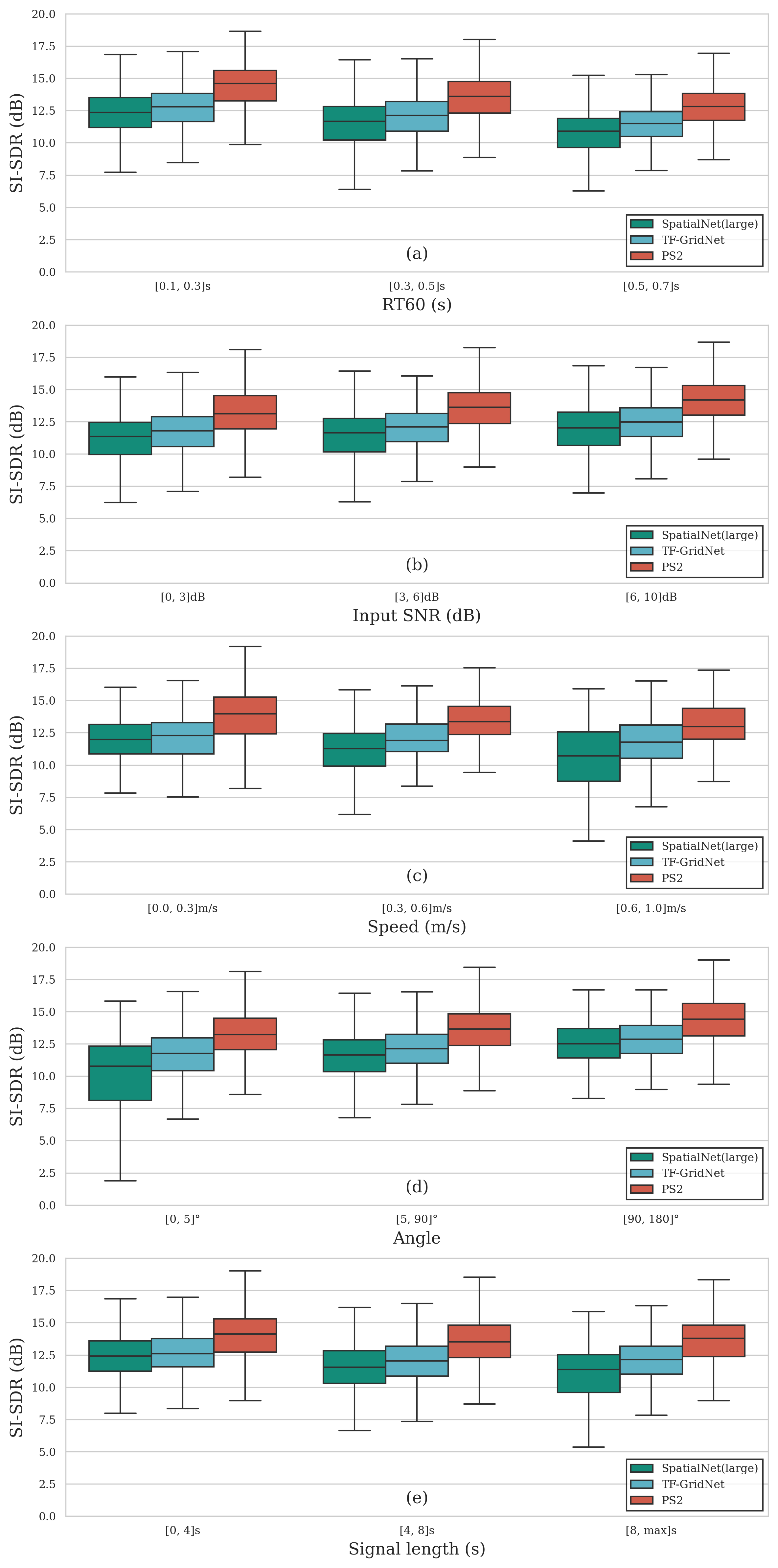}\\ \vspace{-6pt}
  \caption{Multi-conditional evaluation results under different acoustic conditions. SI-SDR performance across varying (a) reverberation times (RT60), (b) input SNR levels, (c) source movement speeds, (d) inter-source angles, and (e) signal durations.
  }
 \vspace{-3ex}
 \label{fig:diagram}
\end{figure}

The test samples are grouped into subsets according to five key acoustic parameters: reverberation time (RT60), input SNR, source movement speed, inter-source angle, and signal duration, as shown in Fig.~\ref{fig:diagram}. To ensure statistical reliability, we maintain a minimum of $500$ samples in each subset. For subsets without sufficient samples, we synthesize additional test samples following the same generation process and acoustic parameters described in Section~\ref{ssec:data}. 
\subsubsection{Separation Performance vs. Reverberation Time}
The impact of reverberation time (RT60) on separation performance was evaluated across three ranges: $[0.1, 0.3]$~s, $[0.3, 0.5]$~s, and $[0.5, 0.7]$~s. PS2 system consistently outperformed other models across all RT60 ranges, achieving a median SI-SDR of $14.6$~dB in low reverberation conditions ($[0.1, 0.3]$~s), which decreased to $12.8$~dB in high reverberation conditions ($[0.5, 0.7]$~s). The performance gap between the PS2 system and other models widened as RT60 increased, with approximately $1.8$~dB SI-SDR advantage over TF-GridNet and $2.2$~dB over SpatialNet in the most challenging test conditions.
The model demonstrated robust performance even in highly reverberant conditions, with the smallest interquartile range (IQR) among all models, indicating more consistent separation quality across different reverberation times.
\begin{table}[t]
\caption{\textsc{Computational Efficiency Comparison}}
\vspace{-1.5ex}
\label{tab:efficiency}
\centering
\begin{tabular*}{\columnwidth}{@{\extracolsep{\fill}}lcccc@{}}
\toprule
Model & Params & FLOPs & GPU Mem & Infer. Time \\
      & (M)    & (G/s)  & (MB/s)  & (ms/s)      \\
\midrule
TF-GridNet          & 14.5 & 463.7 & 533 & 226 \\
SpatialNet(small)   & 1.6  & 46.3  & 145 & 38  \\
SpatialNet(large)   & 7.3  & 237.9 & 193 & 86  \\
PS2                 & 8.4  & 66.8  & 210 & 49  \\
\bottomrule
\end{tabular*}
\vspace{-0.2cm}
\begin{flushleft}
\footnotesize
\textit{Notes}: \\
\yuzhu{
(a) All models evaluated under identical conditions: $4$-second segment, $6$-channel, $16$kHz sampling rate, batch size $1$, NVIDIA RTX 3080. \\
(b) FLOPs are computed using the same method as described in Table~\ref{tab-ablation}, with values reported in Giga per second (G/s). \\
(c) GPU memory is the average memory allocated when processing $4$-second segment over $1000$ runs, divided by $4$ and reported in MB/s. \\
(d) Inference time is the average time to process $4$-second segment over $1000$ runs, divided by $4$ and reported in ms/s.}
\end{flushleft}
\vspace{-0.4cm}
\end{table}
\begin{figure*}[!tbp]
 \centering
 \begin{minipage}{\textwidth}
   \centering
   \includegraphics[width=170mm]{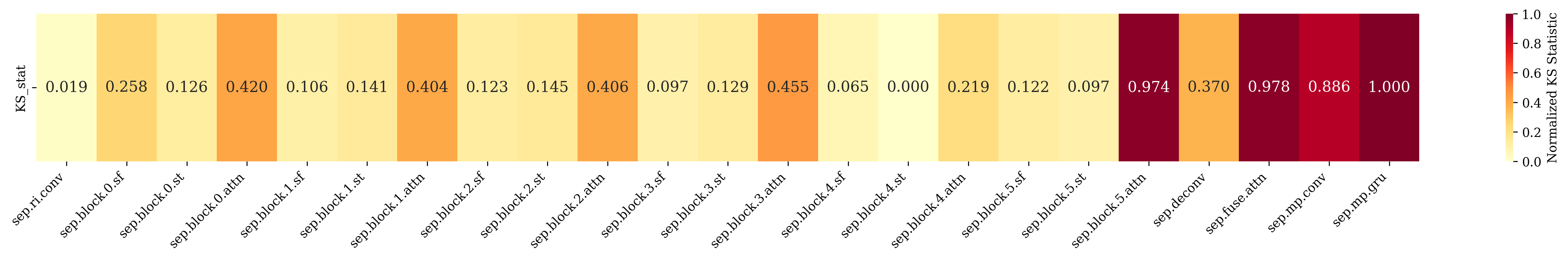}
   \text{(a) PS2}
 \end{minipage}
 \vspace{4pt}
 \begin{minipage}{\textwidth}
   \centering
   \includegraphics[width=170mm]{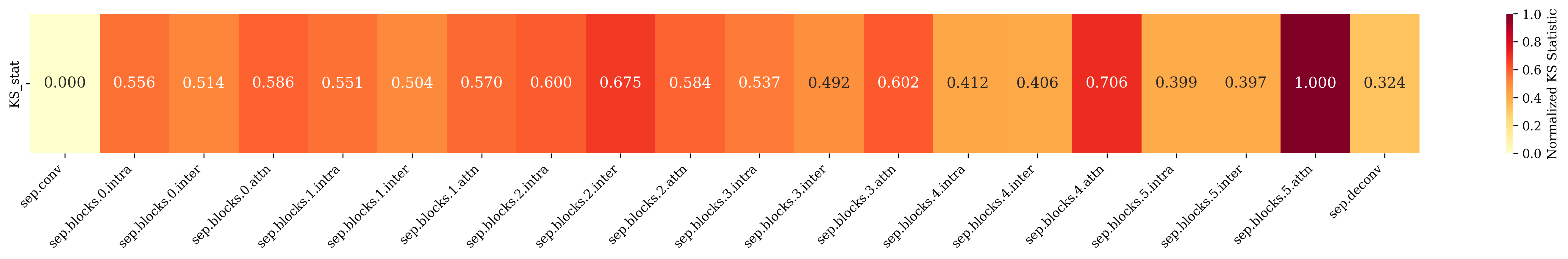}
   \text{(b) TF-GridNet} 
 \end{minipage}
 \vspace{4pt}
 \begin{minipage}{\textwidth}
   \centering
   \includegraphics[width=170mm]{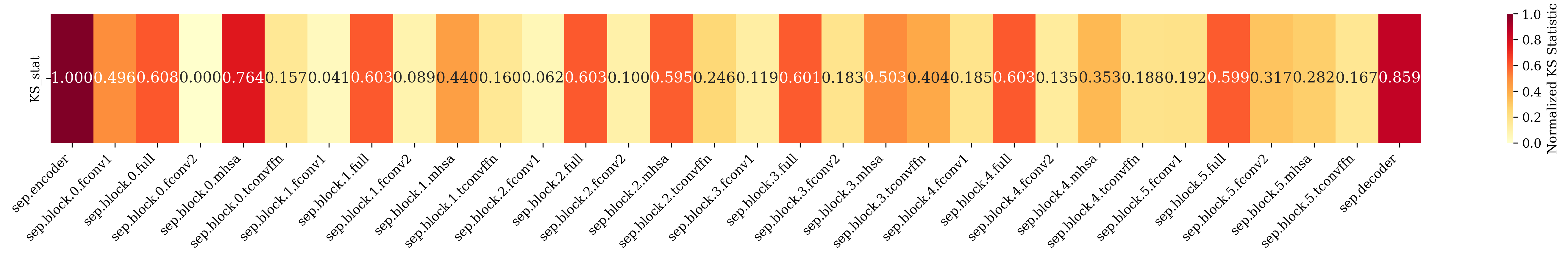}
   \text{(c) SpatialNet}
 \end{minipage}
 \vspace{-1ex}
\caption{Functional block level model sensitivity analysis.
The horizontal axis labels represent functional components.
For (a) PS2 system, 
sep.ri.conv: spectral branch encoder,
sep.block.x.sf: frequency module, 
sep.block.x.st: temporal module, 
sep.block.x.attn: self-attention module in spectral branch, 
sep.deconv: decoder
sep.fuse.attn: cross-attention fuse module,
sep.mp.conv: spatial branch encoder, 
sep.mp.gru: GRU module.
For (b) TF-GridNet, 
sep.conv: encoder, 
sep.block.x.intra: intra-frame full-band module,
sep.block.x.inter: sub-band temporal module,
sep.block.x.attn: cross-frame self-attention module,
sep.deconv: decoder.
For (c) SpatialNet, 
sep.encoder: encoder, 
sep.block.x.fconv1: frequency-convolutional module,
sep.block.x.full: full-band linear module,
sep.block.x.fconv2: frequency-convolutional module,
sep.block.x.mhsa: MHSA module,
sep.block.x.tconvffn: T-ConvFFN module,
sep.decoder: decoder.
}
\vspace{-1ex}
 \label{fig:diagram-sensitivity}
\end{figure*}
\subsubsection{Separation Performance vs. Input SNR}
The evaluation across different input SNR ranges ($[0, 3]$~dB, $[3, 6]$~dB, and $[6, 10]$~dB) revealed that all models benefited from higher SNR conditions. PS2 system exhibited the strongest noise robustness, maintaining a median SI-SDR of $12.5$~dB in the lowest SNR range, which improved to $14.8$~dB in high SNR conditions. The performance advantage was evident in low-SNR scenarios, where the PS2 system outperformed TF-GridNet by approximately $2.0$~dB and SpatialNet by $2.5$~dB in SI-SDR. 
\subsubsection{Separation Performance vs. Movement Speed}
Source movement speed significantly impacted separation performance, with three evaluated ranges: $[0.0, 0.3]$~m/s, $[0.3, 0.6]$~m/s, and $[0.6, 1.0]$~m/s. PS2 system demonstrated superior tracking capability, maintaining median SI-SDR above $13.5$~dB for slow movement and $12.8$~dB for fast movement. The performance degradation with increasing speed was less pronounced for the PS2 system (approximately $0.7$~dB reduction) compared to TF-GridNet ($1.0$~dB) and SpatialNet ($1.2$~dB).
When the movement speed of the sound source is low, the feature changes in the mixture caused by the sound source's movement can be considered as noise. Most of the current high-performing models exhibit a certain level of robustness when dealing with slow-moving sound sources. However, when the sound source moves at a faster speed, its movement can no longer be treated as noise, and the model's ability to handle moving sound sources becomes evident. In scenarios with significant sound source movement, the SI-SDR metric IQRs of the baseline models increase significantly.
\subsubsection{Separation Performance vs. Angle}
In this experiment, we define the inter-source angle as the minimum value of spatial angle (in 3D space) formed by two sound sources at any moment, with the reference microphone as the observation point.
The angles are categorized into three ranges: $[0, 5]^{\circ}$, $[5, 90]^{\circ}$, and $[90, 180]^{\circ}$. All models showed better performance with larger angular conditions. PS2 system achieved the highest median SI-SDR of $14.9$~dB for well-separated sources ($90^{\circ}$-$180^{\circ}$) and maintained $13.2$~dB even for closely spaced sources ($0$-$5^{\circ}$). The performance advantage was most pronounced in challenging small-angle scenarios, where the PS2 system outperformed baselines by $2.3$~dB in SI-SDR.
When the inter-source angle is between $0$ and $5$ degrees, it indicates that the spatial directions of the two sound sources are very close from the perspective of the reference microphone. Under this condition, the SpatialNet shows a significant decline in SI-SDR performance, along with a substantial increase in IQR. In contrast, the PS2 system demonstrates robustness. This indicates that the proposed model exhibits high spatial resolution when tracking moving speakers.
\subsubsection{Separation Performance vs. Signal Length}
The impact of signal duration was analyzed across three ranges: $[0, 4]$~s, $[4, 8]$~s, and $[8, max]$~s, where the maximum signal duration in the test set is $16.1$~s. PS2 system showed consistent performance across different durations, with median SI-SDR values of $14.1$~dB, $13.5$~dB, and $13.8$~dB respectively. The slight performance variation across duration ranges (approximately $0.6$~dB) was smaller than that observed in baseline models ($0.8$-$1.0$~dB), suggesting better temporal modeling capabilities.
Both the PS2 system and TF-GridNet demonstrated robust performance across varying signal lengths, which can be attributed to their effective temporal modeling modules - Mamba and BLSTM respectively. However, SpatialNet, which relies primarily on CNN-based temporal modeling, showed more significant performance degradation for longer sequences, particularly when test samples exceeded the training length. This was evidenced by a notable increase in IQR and a decrease in SI-SDR (dropping by approximately $1.2$~dB) for longer sequences.
The consistent performance across different signal lengths indicates that the proposed PS2 system effectively models both short-term and long-term dependencies during the separation process.
\subsection{Computational Efficiency Analysis}
\yuzhu{We compare the computational efficiency and inference speed of different systems under identical conditions. Table~\ref{tab:efficiency} presents parameter counts, FLOPs, GPU memory usage, and inference time for each system.}

\yuzhu{Compared to TF-GridNet, PS2 uses fewer parameters ($8.4$M vs $14.5$M) with substantially lower FLOPs ($66.8$ vs $463.7$~G/s) and $4.6\times$ faster inference ($49$ vs $226$~ms/s). Compared to SpatialNet(large) with comparable parameters ($7.3$~M vs $8.4$~M) and GPU memory ($193$ vs $210$ MB/s), PS2 achieves substantially lower FLOPs ($66.8$ vs $237.9$~G/s) and $1.8\times$ faster inference ($49$ vs $86$~ms/s). }
\subsection{Sensitivity evaluation}
To evaluate the sensitivity of different models to source movement, we do a sensitivity analysis based on the Kolmogorov-Smirnov (KS) test~\cite{massey1951kolmogorov}.
The same model is trained on two datasets—moving and static—resulting in two trained models, $M_{\text{moving}}$ and $M_{\text{static}}$. 
For each functional block in the model (e.g., frequency module, temporal module in PS2 system), we collect all trainable parameters including weights and biases. The parameters within each functional block are flattened into one-dimensional arrays to collect parameter distribution.
In this process, the order of these parameters in the flattened arrays does not affect the KS statistic calculation.
The KS statistic is computed by comparing the maximum difference (supremum) between the cumulative distribution function (CDF) of two distributions for the $M_{\text{moving}}$ and $M_{\text{static}}$ parameters of the corresponding functional blocks, 
\begin{equation}
    \text{ks} = \sup|F_{\text{moving}}(x) - F_{\text{static}}(x)|,
\end{equation}
where $F_{\text{moving}}(x)$ and $F_{\text{static}}(x)$ are the CDFs.
The KS statistic is normalized to $[0,1]$ range using min-max normalization as the sensitivity measurement, with the minimum and maximum values determined from the KS statistics within each functional block,
\begin{equation}
\text{ks}_{\text{normalized}} = \frac{\text{ks} - \text{ks}_{\text{min}}}{\text{ks}_{\text{max}} - \text{ks}_{\text{min}}}.
\end{equation}
Here, higher values indicate greater distributional differences between static and moving conditions. 
\yuzhu{It should be noted that these differences may reflect the intrinsic sensitivity of modules to source movement, potential differences in how features are weighted during training on different datasets, or a combination of both factors.}
To ensure statistical reliability, we conduct analysis at the functional block level (such as frequency module or temporal module described in Section~\ref{sec:DynamicNet}) rather than at individual layer level, as some layers contain only a few hundred parameters, which could lead to unstable statistical results.
The normalized sensitivity scores are visualized as a heatmap in Fig.~\ref{fig:diagram-sensitivity}, with functional blocks arranged along the x-axis and color intensity representing sensitivity level.
The number of repeated functional blocks in all models is set to six to ensure comparable network depths.
It should be noted that the TF-GridNet model comprises two stacked subnetworks (DNN$_1$ and DNN$_2$) sharing an identical architecture, with only minor differences in input dimensions. Only DNN$_1$ of TF-GridNet is employed in our sensitivity analysis to ensure fair comparison across different models.

As shown in Fig.~\ref{fig:diagram-sensitivity}, the sensitivity patterns reveal different behaviors across different architectures. In the PS2 system, the spatial branch and spectral branch show different sensitivity patterns. The spatial branch shows higher sensitivity in its GRU components (sensitivity $>$ $0.8$) and encoder components (sensitivity $>$ $0.7$), while the spectral branch maintains relatively lower sensitivity (average sensitivity $<$ $0.4$). This different pattern validates our architectural design of separating spectral and spatial processing streams, as it demonstrates that these two types of features indeed undergo different adaptation behaviors when handling moving sources. The attention components in both branches demonstrate moderate to high sensitivity ($0.5$-$0.7$), with slightly higher values in the spatial branch. The fusion module shows intermediate sensitivity (approximately $0.6$), aligning its role in balancing the contributions from both branches based on acoustic conditions.

TF-GridNet shows varied sensitivity across its components, with all attention components showing the highest sensitivity ($>$ $0.8$). This suggests that without explicit branch separation, the model relies heavily on attention mechanisms to handle both spectral and spatial adaptations simultaneously. SpatialNet shows elevated sensitivity across multiple components (encoder, decoder, attention, and full-band linear modules all $>$ $0.6$), indicating a more distributed processing related to source movement.

The consistently high sensitivity of attention components across all three architectures (average sensitivity $>$ $0.7$) validates their importance in handling dynamic scenarios. 
However, the more focused sensitivity pattern in the spatial branch demonstrates the effectiveness of our dual-branch architecture in handling moving sound sources. This design concentrates geometric tracking in dedicated components while preserving stable spectral processing.
\vspace{-1ex}
\section{Conclusions}
\label{sec:conclusions1}
This paper presents PS2, a dual-branch architecture for multi-channel speech separation in dynamic environments. We demonstrate the effectiveness of parallel spectral and spatial processing streams, addressing a fundamental limitation in existing architectures that process these inherently different features through a sequential network stream. 
The ablation studies demonstrate that each component contributes meaningfully to the overall performance, with the dual-branch design providing a $1.5$~dB improvement in SI-SDR compared to a single-branch configuration. In static scenarios, the PS2 system shows superior performance in both multi-channel and single-channel conditions. In moving source scenarios, our model maintains strong separation quality (SI-SDR: $13.4$~dB), outperforming current state-of-the-art methods by $1.6$-$2.1$~dB.

Second, our comprehensive analysis across various acoustic conditions (reverberation times, noise levels, movement speeds, angular separations, and signal durations) demonstrate that our model maintains robust separation performance even under challenging conditions, with consistent SI-SDR improvements of $1.8$-$2.3$~dB over state-of-the-art baselines.

Third, our sensitivity analysis reveals how different architectural components adapt to source movement. The high sensitivity observed in attention mechanisms and spatial branch confirms their crucial role in dynamic scenarios, while the relative stability of spectral processing components aligns with the intuition that speech content features remain invariant to spatial changes.

These findings suggest that the disentanglement of spectral and spatial features is a promising direction for robust multi-channel audio processing in real-world dynamic environments. 
This parallel feature processing approach not only benefits speech separation but also provides insights for other multi-channel audio processing tasks, such as enhancement, dereverberation, and localization, where the different nature of spectral and spatial features has been largely overlooked.

\newpage

\bibliographystyle{IEEEtran}
\bibliography{Bib_YWang}

\end{document}